\documentclass[pra,twocolumn,a4paper,showpacs,floatfix]{revtex4}
\usepackage{epsfig} 
\begin{document}
\title{Fano resonance in discrete lattice models: controlling lineshapes
with impurities}
\author{Arunava Chakrabarti$^\dag$}
 \affiliation{Max-Planck-Institut f\"{u}r Physik komplexer Systeme \\N\"{o}thnitzer Strasse 38,  
Dresden 01187, Germany}
\begin{abstract}
\begin{center}
{\bf Abstract}
\end{center}
\vskip .25in

The possibility of controlling Fano lineshapes in the electronic transmission is  
addressed in terms of a simple discrete model within a tight binding framework, 
in which a finite sized ordered chain is coupled from one side to an infinite 
linear chain (the `backbone') at one lattice point.  
It is found that, the profile of Fano resonance is strongly influenced by  
the presence of impurity atoms in the backbone. 
We specifically discuss the case with just two substitutional impurities sitting 
in the otherwise ordered backbone. Precise analytical formulae relating the locations
of these impurities to the size of the side coupled chain have been presented. The nature of 
the transmission spectrum and the reversal of the pole-zero structures in the Fano resonance
are discussed with the help of these formulae.
\end{abstract}
\maketitle 
\vskip .25in
\noindent
{\bf 1. Introduction}
\vskip .25in
%%%%%%%%%%%%%%%%%%%%%%%%%%%%%%%%%%%%%%%%%%%%%%%%%%%%%%%%%%%%%
The Fano resonance usually refers to a sharp asymmetric profile
in the transmission or absorption spectrum and, is common to many physical 
systems including the absorption of light by atomic systems \cite{fano61}, 
Aharonov-Bohm interferometers \cite{nockel94}-\cite{koba02}, quantum dots \cite{gold00}-\cite{torio04},
propagation of light through photonic crystal waveguides \cite{fan02}-\cite{fan03}.
A Fano profile in the transmission (absorption) 
spectrum is caused by the formation of a discrete energy level in a continuum, and there is 
interaction between the two. A defect present in the system is capable of creating such a 
discrete level. It also 
provides additional paths in the wave scattering \cite{andrey05}. The resulting constructive  
or destructive interference  gives rise to either perfect transmission or complete 
reflection producing a sharp asymmetric lineshape. Fano \cite{fano61} derived this lineshape as,
\begin{equation}
{\cal F}(\epsilon) = \frac{(\epsilon+\eta)^{2}}{\epsilon^{2}+1}
\end{equation}
where, $\epsilon=(E-E_R)/(\Gamma/2)$ with $E_R$ being the `resonance energy' and $\Gamma$ the 
line width. $\eta$ is the asymmetry parameter.

In recent times, it has been demonstrated that, simple discrete lattice 
models with side coupled impurities also exhibit Fano profiles in the transmission spectrum 
\cite{prosen95}-\cite{mirosh05}. Discrete models are important, particularly in the 
context of commendable advancement of miniaturization techniques, as in many cases 
they serve the purpose of understanding some salient features of electronic transport in 
mesoscopic systems. For example, the electronic transmission through a non-interacting 
tight binding chain with an interacting {\it side coupled} quantum dot was examined by Torio et al
\cite{torio02} to investigate the Kondo resonance and Fano anti-resonances in the transport through 
quantum dots. Rodr\'{i}guez et al \cite{rodrig03} modelled a quantum wire attached to a quantum 
dot at one lattice point by a simple discrete tight binding hamiltonian and studied the dynamics 
of electron transport. Fano resonances and Dicke effect \cite{dick53} in quantum wire with side coupled 
quantum dots have been studied, again within a tight binding formalism and independent electron 
approximation, by Guevara et al \cite{guev03} and Orellana et al \cite{orel06}. Interestingly, 
Papadopoulos et al \cite{papa06} have addressed the possibility of controlling electron transport through Fano resonances 
in molecular wires using a first principles approach. Once again,  
the generic features of the transport properties have been captured with the help of a discrete lattice model.

The simplest model which describes the essential features 
of the coupling between a discrete state and a continuum is the Fano-Anderson model \cite{mahan}.
In such a model, one takes a one dimensional ordered lattice and fixes, from one side, a 
`hanging' impurity, which may consist of a single atomic site \cite{prosen95} or a chain with multiple sites
\cite{mirosh05}. The system is described by a tight binding hamiltonian, and 
the role of the `coupling' between the continuum and the discrete levels is  played by
the hopping integral connecting the ordered backbone and the first site of the defect system.
Such a linear chain with a side coupled defect was addressed earlier
\cite{guin87}, but the focus was to understand the effect of such a `geometrical defect'
on the electronic spectrum and localization properties.  
Recently, the transmission profile has been 
examined \cite{arunava06} when the defect chain is a quasiperiodic Fibonacci lattice, which offers 
a multifractal Cantor set spectrum as the chain length becomes infinity. The transmission windows 
have been found to be punctuated by zeros of transmission, the distribution resembling a self similar
fractal as the attached quasiperiodic defect chain becomes larger and larger. 

A very interesting feature in the transmission spectrum of a linear ordered chain with a 
side coupled atom or a chain of atoms \cite{mirosh05}  
is a swapping of the peak-dip (pole-zero) profile of the Fano resonance at special values of the
energy of the electron. In a discrete model, the 
 swapping is found to be associated with the presence of an additional defect (apart from 
the side coupled one) in the ordered
backbone. This additional defect may be a single substitutional {\it impurity} sitting `on' the backbone \cite{mirosh05}, 
or another chain of atoms connected to a different lattice point \cite {arunava06}. 
The important thing here is that, 
the swapping of the peak-dip profile can be controlled by adjusting the number of atoms in the ordered 
backbone which separate the hanging Fano defect chain from the impurity 
\cite{arunava06}.
That is, the control parameter here is a discrete variable
in contrast to previous studies \cite{kim99}-\cite{ladron06} where one had to vary continuous parameters 
such as the strength of the impurity potential or the magnetic field to see such an effect. 
The significance of such swapping has already 
been appreciated in the literature \cite{voo05}. For example, if 
similar reversal of peak position happens even with a magnetic impurity, then the transmission 
peak of the spin `up' (down) electron may coincide with the transmission zero of the spin `down' (up) 
electron, making the system behave as a {\it spin filter}. Incidentally, a tight binding model of a 
quantum wire with an attached ring has been proposed as a model of a spin filter 
\cite{lee06} and another discrete $T$-stub structure has very recently been examined as a potential 
candidate for a spin-splitting device \cite{wang06}.

In this context, the possibility of controlling the Fano profiles by placing impurities on the 
backbone becomes an interesting issue. 
This important aspect has received very little attention, specially in the 
context of a discrete lattice model.
This precisely is the idea which motivates us to 
look deeper into the problem of tuning Fano lineshapes in a discrete lattice model. In this
communication, we present the 
first step taken by us to look into the effect of placing more than one impurity 
in the ordered backbone
on the transmission lineshapes. Secondly, we examine carefully whether there is a correlation 
between the `size' of the side coupled Fano-Anderson (FA) defect and the separation between the impurities 
on the ordered backbone. These aspects may turn out to be important in gaining control
over the Fano profiles in the transmission, particularly, over a possible reversal of the peak-dip
profile.  

We get extremely interesting results. In this paper we present analytical results with just two impurities 
sitting symmetrically on the backbone $N$ sites away from the lattice point at which the defect chain 
of $n$ sites is side attached (Fig.1).
We show, that there is a definite correlation between the numbers $N$ and $n$ which controls the 
transmission spectrum in a {\it completely predictable} way. 
This is also intimately connected to the signs of the impurity potentials.
We provide a prescription to predict the 
principal features of the transmission spectrum and the possibility of securing a full control 
over the reversal of the 
peak-dip structures. This aspect, to the best of our knowledge, has not been addressed before.  
In what follows, we describe the model and the method in section 2. Sections  3 and 4 contain the results and 
the discussion and, in section 5 we draw conclusions.
%%%%%%%%%%%%%%%%%%%%%%%%%%%%%%%%% FIGURE 1 IS BELOW %%%%%%%%%%%%%%%%%
\begin{center}
\begin{figure}
%\centering \figspace
{\centering \resizebox* {7cm}{4cm}{\includegraphics{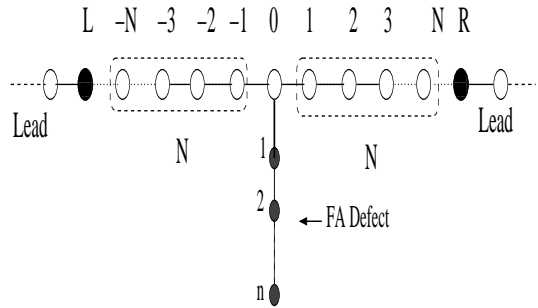}}}
%\centerline{\includegraphics{fig1.eps}}
\caption{\label{fig1} A Fano-Anderson (FA) 
defect coupled to a discrete chain (the backbone) at one site (marked `zero').
The two `impurities' (`$L$' and `$R$') are marked by solid circles on the backbone. }
\end{figure}
\end{center}
%%%%%%%%%%%%%%%%%%%%%%%%%%%%%%%%%%%%%%%%%%%%%%%%%%%%%%%%%%%%%%%%%%%%
%%%%%%%%%%%%%%%%%%%%%%%%%%%%%%%%%%%%%%%%%%%%%%%%%%%%%%%%%%%%%
\vskip .25in
\noindent
{\bf 2. The Model and the method}
\vskip .25in
Let us refer to Fig.1. The backbone (model quantum wire) consists of periodically
placed identical sites (open circles) excepting two `impurities' (solid black circle), each 
being separated from the zeroth site by $N$ sites of the 
ordered chain. At the zeroth site of the backbone 
a $n$-site Fano-Anderson (FA) chain is attached, which we shall refer to as the FA defect. 
Beyond the impurities (black circles on the backbone), the 
remaining portions of the backbone serve as the `leads' along which an electron enters the system, 
and leaves it as well. The FA defect plays the role of a resonant cavity sustaining standing waves \cite{prosen95}.
In one of the first papers on similar problems, the influence of a grafted stub in a quantum wire
in obtaining Fano resonance was discussed earlier, using 
the continuous version of the Schr\"{o}dinger equation by Tekman and Bagwell \cite{bag93}.

The Hamiltonian of the system, in the standard tight binding form, is written as, 
\begin{equation}
H = H_{W} + H_{D} + H_{WD}
\end{equation}
where, 
%%%%%%%%%%%%%%%%%%%%%%%%%%%%%%%%
\begin{eqnarray}
H_{W} & = & \sum_{i=-\infty}^{\infty} \epsilon_i c_i^{\dag} c_i + t_0 \sum_{<ij>} c_i^{\dag} c_j \nonumber \\
H_{D} & = & \sum_{i=1}^{n} \epsilon_i d_i^{\dag} d_i +  
\sum_{<ij>} t_{ij} d_i^{\dag} d_j \nonumber \\
H_{WD} & = & \lambda (c_0^{\dag}d_1 + d_1^{\dag}c_0) 
\end{eqnarray}
In the above, $c^{\dag}(c)$ and $d^{\dag}(d)$ represent the 
creation (annihilation) operators for the wire and the defect respectively.
The on-site potentials on the ordered backbone are, $\epsilon_i = \mu_L$ for the left (L) impurity, 
and $\epsilon_i = \mu_R$ for the right (R) impurity, and 
$\epsilon_i = \epsilon_0$ for the rest of the chain, as well as for every site of the FA defect.
$t_0$ is the constant hopping integral in the 
periodic array of sites, which is our `wire'.  
For all the results in this paper we assign the same hopping integral $t_0$ among the sites in the bulk of the 
FA defect. The first site of the FA defect is coupled to the zeroth site of the backbone via a hopping integral
$\lambda$.
So, $\lambda$ is the `interaction'
which locally couples the two subsystems, viz, the periodic chain, and the dangling defect. 
If we decouple the two, the spectrum of the periodic quantum wire is
absolutely continuous, the band extending from $\epsilon_0-2t_0$ to $\epsilon_0+2t_0$. The dispersion
relation is $E=\epsilon_0+2t_0 \cos qa$, $a$ representing the lattice spacing. With non-zero $\lambda$ it is 
not right to view the two subsystems separately. In this case, the defect chain may be looked upon as a 
single impurity with a complicated internal structure and located at a single site marked `zero' 
in the backbone (Fig.1).
Therefore, while examining the transmission across 
the impurity, one has to be careful to adjudge whether the concerned energy really belongs to the 
spectrum of the entire system, that is, the ordered backbone plus the FA defect.

To calculate the electronic transmission across such a system, we first of all `wrap' the hanging
FA defect to create an effective defect site with an energy dependent on site potential $\epsilon^*$
at the lattice point marked zero. This is easily done
by making use of the set of difference equations
\begin{eqnarray}
(E-\epsilon_0) \psi_1 & = & \lambda \psi_{0(Backbone)} + t_0 \psi_2 \nonumber \\
(E-\epsilon_0) \psi_j & = & t_0 (\psi_{j-1} + \psi_{j+1}) \nonumber \\
(E-\epsilon_n) \psi_n & = & t_0 \psi_{n-1}
\end{eqnarray}
for all sites of the FA defect chain with $2 \le j \le n-1$ corresponding to the middle equation, 
and then eliminating all the individual amplitudes $\psi_j$ with $j=1$ to $n$ 
sequentially in terms 
of the site number zero in the backbone. Using this decimation renormalization method \cite{arunava06},
\cite{south83},
the effective (renormalized) on site potential $\epsilon^*$ at the zeroth site of the backbone
is given by, 
\begin{equation}
\epsilon^* = \epsilon_0 + \frac{\lambda^2}{E-\tilde\epsilon}
\end{equation}
where, 
\begin{equation}
\tilde\epsilon = \epsilon_0 + \frac{t_0U_{n-3}(x)}{U_{n-2}(x)}+\frac{t_0^2}{U_{n-2}^2(x)}
\frac{1}{E-\epsilon_0-\frac{t_0U_{n-3}(x)}{U_{n-2}(x)}}
\end{equation}
for $n \ge 2$.
Here, $x=(E-\epsilon_0)/2t_0$ and $U_n(x)$ is the $n$th order Chebyshev polynomial of the second kind.
The problem is now reduced to the calculation of transmission across a linear array of three impurities 
with on-site potentials $\mu_L$, $\epsilon^*$ and $\mu_R$ occupying lattice coordinates $-N$, $0$ and 
$N$ respectively. The result is,
\begin{equation}
T = |\tau^2|
\end{equation}
with the transmission `amplitude' $\tau$ given by \cite{stone81}, 
\begin{equation}
\tau = e^{-i\theta(q,N)} \frac{2 \sin qa}{\alpha + i\beta}
\end{equation}
Here,
\begin{eqnarray}
\alpha & = & (M_{11} - M_{12}) \sin qa - M_{22} \sin 2qa \nonumber \\
\beta & = & M_{11} + (M_{12} - M_{21}) \cos qa - M_{22} \cos 2qa
\end{eqnarray}
and, 
$q$ and $a$ are the wave vector and the lattice spacing respectively. We shall take $a=1$ 
in all the calculations which follow. $\theta(q,N)$ is a phase factor, whose explicit form is not displayed as it will not
be needed for our purpose.
$M_{ij}$ are the elements of the transfer matrix, 
\begin{equation}
M = M_R.M_{Cent}.M_L
\end{equation}
where,
\begin{equation}
M_{R(L)} =  \left( \begin{array}{cc}\frac{E-\mu_{R(L)}}{t_0} & -1 \\1 & 0 \end{array} \right ).
\end{equation}
and, 
\begin{widetext}
\begin{equation}
M_{Cent} =
  \left( \begin{array}{cc}U_N(x) & -U_{N-1}(x) \\U_{N-1}(x) & -U_{N-2}(x) \end{array} \right ).
  \left( \begin{array}{cc}(E-\epsilon^*)/t_0 & -1 \\1 & 0 \end{array} \right ).
  \left( \begin{array}{cc}U_N(x) & -U_{N-1}(x) \\U_{N-1}(x) & -U_{N-2}(x) \end{array} \right ).
\end{equation}
\end{widetext}
 
%%%%%%%%%%%%%%%%%%%%%%%%%%% FIGURE 2 IS HERE %%%%%%%%%%%%%%%%%%%%%%%%%%%%%%%%%
\begin{center}
\begin{figure}
%\centering \figspace
{\centering \resizebox* {16cm}{14cm}{\includegraphics[angle=-90]{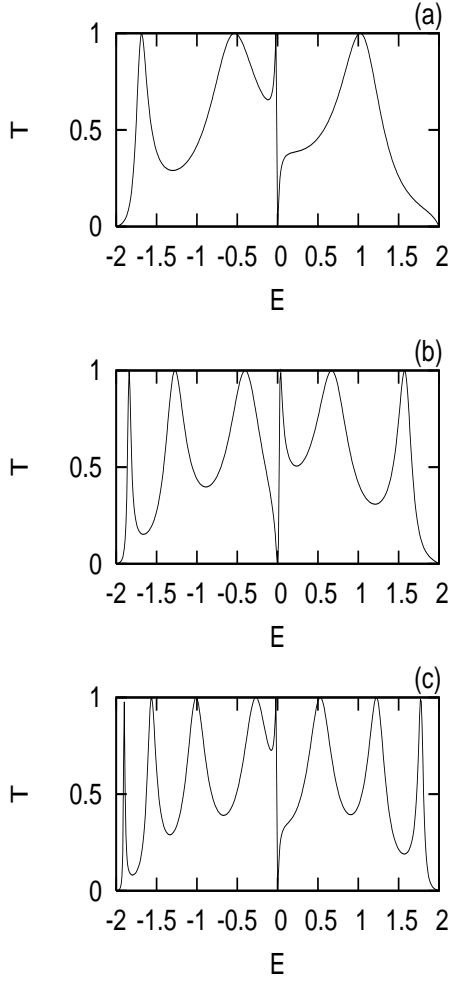}}}
%\centerline{\includegraphics[width=0.95\columnwidth,angle=-90]{fig2.eps}}
\caption{\label{fig2}The transmission coefficient for single atom FA defect. The 
two impurities on the backbone have identical signs, viz, $\mu=+1$ for each of them.
We have taken $\epsilon_0=0$, $t_0=1$ and $\lambda=0.2$ in unit of $t_0$. The peak-dip
profile at $E=\epsilon_0=0$ in (a) ($N=1$) is swapped into a dip-peak structure in (b) ($N=2$), and 
is reversed in (c) ($N=3$).}
\end{figure}
\end{center}
%%%%%%%%%%%%%%%%%%%%%%%%%%%%%%%%%%%%%%%%%%%%%%%%%
\noindent
{\bf 3. Results and discussion}
\vskip .25in
The general features of the transmission spectra for various values of $N$ and $n$ are
presented in Fig.2 to Fig.5.
To justify the spirit of this work, we discuss specifically Fano lineshape in the transmission 
and their reversal at special 
energies.
The influence of the signs of the impurity potentials $\mu_{L(R)}$ are also addressed. 
To simplify matters, we set $|\mu_L| = |\mu_R| = \mu$, and 
to have a clear understanding of the arguments we present 
analytical results when the FA defect is just a single atom. The extension to the cases of
multi-atom FA defect is trivial, only difference is that then we have more than 
one energy at which the reversal of the peak-dip profile takes place.  
\vskip .3in
\noindent
%%%%%%%%%%%%%%%%%%%%%%%%%%%%%%% FIGURE 3 IS HERE %%%%%%%%%%%%%%%%%%%%%%%%%%%%%%
\begin{figure}
%\centering \figspace
{\centering \resizebox* {16cm}{14cm}{\includegraphics[angle=-90]{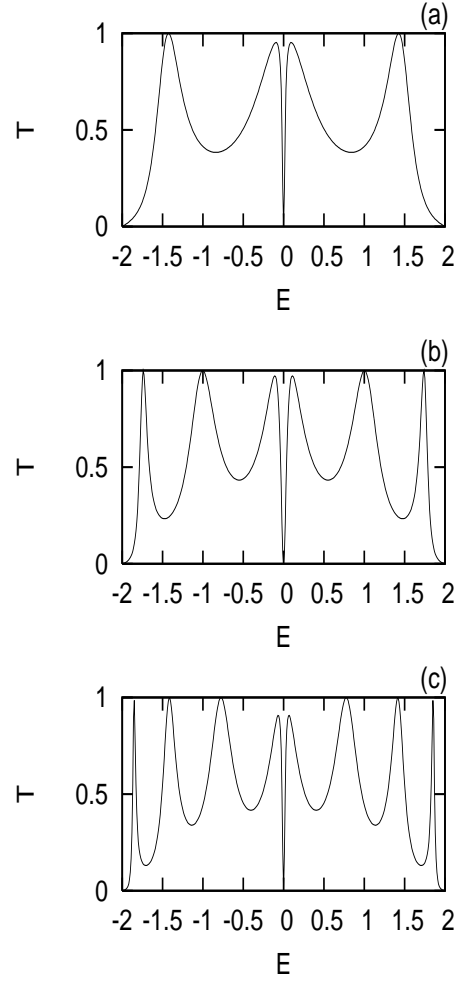}}\par}
%\centerline{\includegraphics[width=0.95\columnwidth,angle=-90]{fig3.eps}}
\caption{\label{fig3}The transmission coefficient ($T$) 
against energy for single atom FA defect when the impurity-potentials on the 
backbone are set equal to $\mu=+1$ and $\mu=-1$ for the left (L) and the right (R)
impurities respectively. Other parameters are the same as in Fig.2.}
\end{figure}
%%%%%%%%%%%%%%%%%%%%%%%%%%%%%%%%%%%%%%%%%%%%%%%%%%%%%%%%%%%

\noindent
{\it 3.1 The repulsive impurity:} $\mu > 0$
\vskip .25in
With just one atom as FA defect ($n=1$), at $E=\epsilon_0$, we get $\epsilon^*=\infty$ resulting in  $T=0$, an 
anti-resonance. The spectrum for $N=1$, $2$ and $3$, has been shown in Fig.2. The peak-dip (pole-zero) 
structure in the spectrum and its reversal with $N$ is clear at the anti-resonance energy. 
To get an analytical insight into the effect of changing $N$,
we expand $\tau$ in Eq.(8) around $E = \epsilon_0$ by setting $E = \epsilon_0+\delta$, and  
drop terms smaller than those of the order of $\delta$. The result, apart from the trivial phase factor in Eq.(8), is,
\begin{equation}
\tau \simeq \frac{(2t_0^3 \sin qa) \delta}{t_0 {\cal F}(N)\left [\delta + 
\frac{{\cal G}(N)}{{\cal F}(N)} \right ]
+ i{\cal H}(N) \left [\delta + \frac{{\cal K}(N)}{{\cal H}(N)} \right ]}
\end{equation}
where, ${\cal F}(N)=\lambda^2U_N^2-2t_0^2(U_{N-1}^2+U_NU_{N-2})+4t_0\Delta U_NU_{N-1}$, 
${\cal G}(N)=2\lambda^2U_N(\Delta U_N-t_0U_{N-1})$, 
${\cal H}(N)=\lambda^2t_0U_NU_{N-1}+2t_0^3U_{N-1}(U_N-U_{N-2})
-\Delta \lambda^2 U_N^2+2t_0^2\Delta(U_{N-1}^2+U_NU_{N-2})-2t_0\Delta^2 U_NU_{N-1}$, and 
${\cal K}(N)=\lambda^2[t_0^2(U_N^2-U_{N-1}^2)-\Delta^2 U_N^2+2t_0\Delta U_NU_{N-1}]$.
We have set $\Delta = \epsilon_0  - \mu$.
All the quantities ${\cal F}$, ${\cal G}$, ${\cal H}$, and ${\cal K}$, and the wave vector $q$ have to be evaluated at the 
$\delta$-neighbourhood of $E=\epsilon_0$, and Fig.2 corresponds to the case where $\epsilon_0=0$ and $\mu=+1$.

The expression of the transmission amplitude reveals that the zero of transmission  
is at $\delta =0$, while the real part in the denominator becomes zero 
when $\delta = -{\cal G}(N)/{\cal F}(N)$. Thus the locations of the two zeros are
different. Similar observations were made earlier in the case of a 
mesoscopic two-lead ring \cite{voo05}. These detuned but concomitant zeros 
result in a non zero asymmetry parameter and an asymmetric Fano lineshape \cite{swarn04}
at $E = \epsilon_0$. The interesting fact about the ratio ${\cal G}(N)/{\cal F}(N)$ is that, it flips sign 
as $N$ changes. This results in a swapping of the peak-dip profile. For example, when 
${\cal G}(N)/{\cal F}(N) < 0$, then the peak of the transmission follows the dip. On the other hand, for 
${\cal G}(N)/{\cal F}(N) > 0$, the peak precedes the dip. This precisely happens in Fig.2, where the 
peak$\rightarrow$dip structure for $N=1$ is reversed into a dip$\rightarrow$peak one 
for $N=2$ at $E=\epsilon_0$. The profile of Fano resonance is of course restored as $N$
is increased to the value $3$. The sign-flip continues with $N$ increasing sequentially (an odd-even effect
in terms of $N$), as can be checked by calculating ${\cal G}(N)/{\cal F}(N)$ for successive values 
of $N$, with $n=1$.
It is also apparent that, apart from the sharp Fano resonance 
at $E=\epsilon_0$, there will be other transmission peaks arising out of the typical Fabry-Perot
type interference effects as the incoming wave interferes constructively in between the impurity sites.

With $\mu=-1$, the entire spectrum is a mirror image (with respect to the centre 
of the spectrum) of Fig.2. So, we move on to 
the next non-trivial case.

\vskip .25in
\noindent
{\it 3.2 One attractive and one repulsive impurity:} $\mu_L=-\mu$ and $\mu_R=+\mu$
\vskip .25in

We now discuss the case when the on-site potentials of the left and the right impurities are 
equal in magnitude, but opposite in sign. That is, $\Delta$ has the same magnitude for the
left and right impurities. Only the sign of $\Delta$ is opposite for them. 
In this case, an expansion of the transmission 
amplitude around $E = \epsilon_0$ as done earlier leads to the result (again dropping the unimportant phase factor), 
\begin{equation}
\tau \simeq \frac{(2t_0^3 \sin qa) \delta}{t_0\Theta(N) \left [\delta - \frac{\Phi(N)}{\Theta(N)} \right ]+
i \left [\delta\Gamma_1 + \Gamma_2 \right ]}
\end{equation}
where, $\Theta(N)=\lambda^2U_N^2-2t_0^2(U_{N-1}^2+U_NU_{N-2})$, $\Phi=2t_0\lambda^2U_NU_{N-1}$, 
$\Gamma_1=2t_0\Delta^2U_NU_{N-1}+2t_0^3U_{N-1}(U_N-U_{N-2})+\lambda^2t_0U_NU_{N-1}$ and, 
$\Gamma_2=\lambda^2[\Delta^2U_N^2+t_0^2(U_N^2-U_{N-1}^2)]$. It is obvious that the zero of 
transmission occurs again at $\delta=0$, while the possibility of a peak-dip reversal is there if  
the ratio $\Phi(N)/\Theta(N)$ flips sign as $N$ changes.
Here, in this case, it can be verified that the ratio $\Phi(N)/\Theta(N)$ {\it does not} change sign
as $N$ increases. This implies, a swapping of the peak-dip profile of the Fano 
resonance at $E=\epsilon_0$ is {\it not possible} in this case. 
The peak always remains on one side of the dip for a fixed sign of $\delta$.
In addition to this, it is to 
be appreciated that the ratio $\Phi(N)/\Theta(N)$ is independent of $\Delta$, and  
yields identical values for $E-\epsilon_0=\pm \delta$. This explains the completely 
symmetrical pattern in the transmission coefficient immediately around $E=\epsilon_0$ in Fig.3. In fact, the 
entire spectrum in this case is symmetric around the centre, i.e. around $E=\epsilon_0$, a fact that 
of course, is due to the same choice of the value of the on-site potential for the FA defect and the 
backbone (except the impurities).

%\section{More atoms in the FA defect: search for a correlation between $n$ and $N$}

\vskip .3in
\noindent
{\bf 4. More atoms in the FA defect: correlation between `$N$' and `$n$'}
\vskip .25in

We now turn to the cases when one has got more atoms in the  FA defect.
The basic question we ask is that, whether the swapping of the zero-pole structure in the Fano profile
remains a generic feature of the transmission, independent of $N$ and $n$ . Now we shall have $n$ eigenvalues for 
an $n$-site FA defect at which the transmission coefficient vanishes. One has to do the exercise 
outlined in the previous section for each of these eigenvalues, which is not difficult, but tiring.
Instead, we present results of systematic evaluation of $T$ for several such cases, from which we have been 
able to work out the precise criterion for the existence of a reversal of the 
sharp asymmetry in the peak-dip profile.
The formulae have been numerically checked by changing $N$ and $n$ at our will and 
have been verified analytically for small 
values of $N$ and $n$. The formulae are found consistent with the transmission spectra   
for all $(N,n)$ combinations we examined.

%%%%%%%%%%%%%%%%%%%%%%%%%%%%%% FIGURE 4 IS HERE %%%%%%%%%%%%%%%%%%%%%%%%%%%%%%
\begin{figure}
%\centering \figspace
{\centering \resizebox* {16cm}{14cm}{\includegraphics[angle=-90]{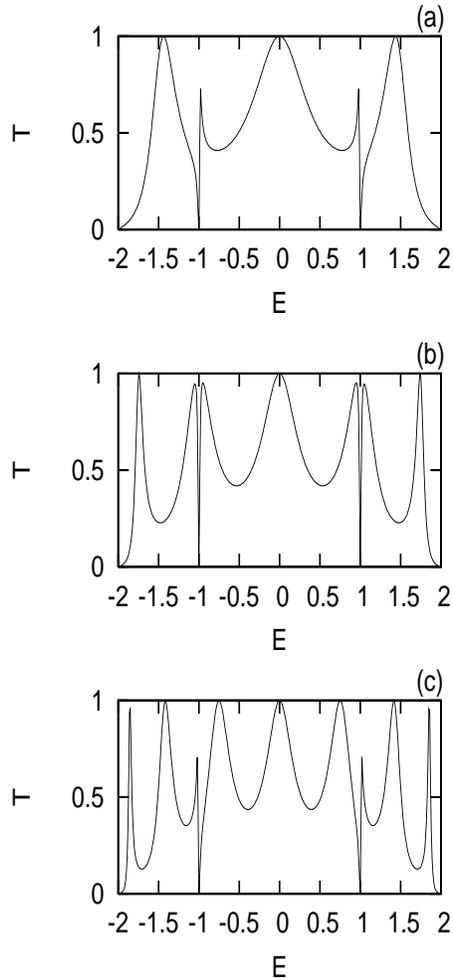}}}
%\centerline{\includegraphics[width=0.95\columnwidth,angle=-90]{fig4.eps}}
\caption{\label{fig4}The swapping of peak-dip profile and the $N$-dependence of transmission
when the side coupled defect chain contains two atoms ($n=2$). Here, $\mu_L=1$ and $\mu_R=-1$ for the 
left and the right impurity respectively. $N=1$, $2$ and $3$ in figures (a), (b) and (c) respectively.
The values of on-site potentials and the hopping integrals are as in Fig.3, and $\lambda=0.2$ in
unit of $t_0$.} 
\end{figure}

%%%%%%%%%%%%%%%%%%%%%%%%%%%%%% FIGURE 5 IS HERE %%%%%%%%%%%%%%%%%%%%%%%%%%%%%%
\begin{figure}
%\centering \figspace
{\centering \resizebox* {16cm}{14cm}{\includegraphics[angle=-90]{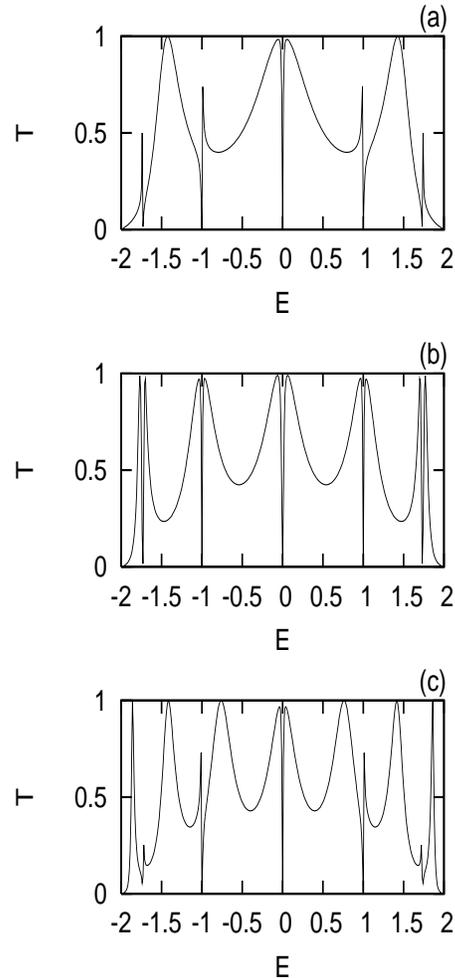}}}
%\centerline{\includegraphics[width=0.95\columnwidth,angle=-90]{fig5.eps}}
\caption{\label{fig4}The swapping of peak-dip profile and the $N$-dependence of transmission
when the side coupled defect chain contains five atoms ($n=5$). 
$N=1$, $2$, $3$ in figures (a), (b) and (c) respectively, and the other parameters are same as in Fig.4.}
\end{figure}

We begin with $n=2$ and only discuss the case when the left and the right impurities
have the same magnitude, but opposite sign. That is, we have a repulsive and an attractive 
potential sitting as impurities on the backbone. We also 
set the FA defect-wire coupling at a relatively small value to make the 
asymmetric jumps in the spectrum sharp. The two atoms in the FA defect have
the same on site potential $\epsilon_0$ and the hopping integral is $t_0$ everywhere. 
While in all cases the spectrum is mirror symmetric around $E=\epsilon_0$ with a transmission zero 
showing up at the centre of the spectrum, we observe remarkable changes at other parts of the spectrum
as $N$ is changed sequentially. This is illustrated in Fig.4 for a weak 
FA defect-wire coupling $\lambda$. For $N=1$, two sharp asymmetric 
Fano resonances are observed at $E=\epsilon_0 \pm t_0$ which are the eigenvalues of an isolated 
two-atom chain. A dip-peak structure appears at $E=\epsilon_0-t_0$, while a peak-dip profile is 
seen at $E=\epsilon_0+t_0$. With $N=2$, the 
asymmetric Fano lineshapes at $E=\epsilon_0 \pm t_0$ {\it totally disappear} and are replaced by {\it almost} 
symmetric anti-resonances. 
We use the word `almost' because with smaller values of $\lambda$, the transmission anti-resonances 
become more and more symmetric looking, while with increasing $\lambda$, the width of the anti-resonance increases, and the 
Fano anti-resonance deviates from a symmetric appearance as we move away from the special energies.  
The overall mirror symmetry 
of the spectrum persists. By changing $N$ from $2$ to $3$ we restore the asymmetric structure in the
transmission spectrum around $E=\epsilon_0 \pm t_0$, but now, remarkably, the sharp peak-dip profiles 
are found swapped with respect to the $N=1$ case. In this way, with $n=2$, we have checked that there 
is {\it no swapping of the transmission profile} (this is important to appreciate) if
we set $N=2$, $5$, $8$ etc, though
the {\it symmetric} look of the transmission anti-resonances are controlled by the 
strength of the coupling $\lambda$. Reversal of the peak-dip profile
occurs for other values of $N$. 
In Fig.5 we present results for $n=5$. At $E=\epsilon_0 \pm t_0$ and $E=\epsilon_0 \pm \sqrt{3}t_0$, we find sharp 
asymmetric Fano lineshapes for $N=1$ which disappear as $N$ changes to $2$. 
With $N=3$ the sharp (local) asymmetry
in lineshapes around the above energies is back, but now the peak-dip sequence is swapped. 
Multiple resonances appear in between.  
The jump at the extreme values like 
$E=\epsilon \pm \sqrt{3}t_0$ appears smaller, but the (local) asymmetry and the sharpness are apparent.
We have carried out extensive
numerical evaluation of the transmission spectrum by changing $n$ and $N$ and for the case when
the left and the right impurity potentials  have same magnitude, but different signs. The results 
can be summarized in two different formulae which have been verified with arbitrary 
combination of $N$ and $n$. The results are the following. 
\vskip .25in

(a) For even number of atoms $n$ in the FA defect, the transmission spectrum will exhibit 
{\it no asymmetric peak-dip (or, dip-peak) profile} when we select $N$ in such a way that, 
\begin{equation}
N_l = (n+1)l - 1
\end{equation}
with $n=2$, $4$, $6$, etc., and $l=1$, $2$, $3$, $....$

\vskip .25in
(b) When $n$ is odd, the formula is, 
\begin{equation}
N_l = \left (\frac{n+1}{2}\right )l -1
\end{equation}
for $n \ge 3$ and $l$ again being equal to $1$, $2$, $3$, etc.
\vskip .3in
Before ending this section, we would like to mention that we have also 
investigated other combinations of $(N,n)$ and the their interplay with the 
signs and the magnitudes of the impurity potential $\mu$. Results indicate the presence of both 
symmetric and asymmetric Fano lineshapes in the transmission spectrum in general.
The reversal of peak-dip structure takes place as $N$ changes sequentially. For example, with 
$\mu_L=\mu_R=\mu > 0$, and $n=2$, the spectrum consists of a sharp asymmetric Fano anti-resonance with a 
dip-peak (zero-pole) jump at $E=\epsilon_0-t_0$ and a symmetric transmission zero at $E=\epsilon+t_0$
if we start with $N=1$. With $N=2$, the spectrum exhibits only asymmetric profiles at the two above energies, 
with the earlier dip-peak shape at $E=\epsilon-t_0$ being swapped into a peak-dip structure. The 
symmetric lineshape at $E=\epsilon_0+t_0$ is now replaced by an asymmetric peak-dip drop. The spectrum
is not symmetric on the whole around the centre, i.e. around $E=\epsilon_0$. Such variations have been 
studied with supporting analytical formulae for various combinations of $N$, $n$ and $\mu$. However, 
we do not show every such case separately to save space.

\vskip .3in
\noindent
{\bf 5. Conclusions}
\vskip .25in
In conclusion, we have examined the occurrence of Fano lineshapes in a discrete model 
within a tight binding framework using analytical and numerical methods. We have shown that 
by placing more than one substitutional impurity in the backbone lattice we can control the 
Fano lineshapes in a completely predictable way. We provide exact formulae which guide us
to choose appropriate combination of the placement of the impurities and the size of the 
side coupled Fano defect for which we either retain, or get rid of the 
swapping of the asymmetric Fano profiles
in the transmission spectrum. Two specific cases have been presented to explain our idea. This 
simple model may help in choosing an appropriate combination of the defect's size and the 
impurity-locations in a quantum wire to gain control over resonance lineshapes. 
It would be interesting to see if similar lineshapes are observed if one incorporates spin in such
discrete models, in which case observing a `spin filtering' effect in such discrete models may 
not be a remote possibility. Work in this direction is currently in progress.
\vskip .35in
\noindent
{\bf Acknowledgments}
\vskip .25in
I am grateful to the Max-Planck-Institute f\"{u}r Physik komplexer Systeme in Dresden
for their hospitality and financial support during the course of this work. I 
acknowledge discussions with Sergej Flach and thank him for bringing 
several important references to my notice.   
I am also thankful to Ricardo Pinto for a careful reading of the manuscript and many constructive criticisms. 
\vskip .3in
\noindent
{\bf References}
\vskip .25in
$^\dag$ On leave from Department of Physics, University of Kalyani, Kalyani, 
West Bengal 741 235, India. E-Mail: arunava@klyuniv.ernet.in
\begin{itemize}
\bibitem{fano61} U. Fano, Phys. Rev. 124 (1961) 1866. 
\bibitem{nockel94} J. U. N\"{o}ckel, A. D. Stone, Phys. Rev. B 50 (1994) 17415. 
\bibitem{koba02} K. Kobayashi, H. Aikawa, S. Katsumoto, Y. Iye, Phys. Rev. Lett. 88 (2002) 256806. 
\bibitem{gold00} J. G\"{o}res, D. Goldhaber-Gordon, S. Heemeyer, M. A. Kastner, H. Shtrikman, 
D. Mahalu, U. Meriav, Phys. Rev. B 62 (2000) 2188.
\bibitem{bul01} B. R. Bulka, P. Stefanski, Phys. Rev. Lett. 86 (2001) 5128.
\bibitem{torio04} M. E. Torio, K. Hallberg, S. Flach, A. E. Miroshnichenko, M. Titov, 
Eur. Phys. J. B 37 (2004) 399.
\bibitem{fan02} S. Fan, J. D. Joannopoulos, Phys. Rev. B 65 (2002) 235112.
\bibitem{yanik03} M. F. Yanik, S. H. Fan, M. Soljacic, Appl. Phys. Lett. 83 (2003) 2739.
\bibitem{fan03} S. H. Fan, W. Suh, J. D. Joannopoulos, J. Opt. Soc. Am. B 20 (2003) 569.
\bibitem{andrey05} A. E. Miroshnichenko, S. F. Mingaleev, S. Flach, Y. S. Kivshar, Phys. Rev. E 71 (2005) 
036626.
\bibitem{prosen95} P. Singha Deo, C. Basu, Phys. Rev. B 52 (1995) 10685.
\bibitem{mirosh05} A. E. Miroshnichenko, Y. S. Kivshar, Phys. Rev. E 72 (2005) 056611.
\bibitem{torio02} M. E. Torio, K. Hallberg, A. H. Ceccatto, C. R. Proetto, 
Phys. Rev. B 65 (2002) 085302.
\bibitem{rodrig03} A. Rodr\'{i}guez, F. Dom\'{i}nguez-adame, I. G\'{o}mez, P. A. Orellana, 
Phys. Lett. A 320 (2003) 242.
\bibitem{dick53} R. H. Dicke, Phys. Rev. 89 (1953) 472.
\bibitem{guev03} M. L. Ladr\'{o}n de Guevara, F. Claro, P. A. Orellana, Phys. Rev. B 67 (2003) 195335.
\bibitem{orel06} P. A. Orellana, F. Dom\'{i}nguez-Adame, E. diez, arXiv:cond-mat/0607094.
\bibitem{papa06} T. A. Papadopoulos, I. M. Grace, C. J. Lambert, arXiv:cond-mat/0609109.
\bibitem{mahan} G. D. Mahan, Many Particle Physics (1993) (New York, Plenum Press).
\bibitem{guin87} F. Guinea, J. A. Verge\'{e}s, Phys. Rev. B 35 (1987) 979.
\bibitem{arunava06} A. Chakrabarti, To be published in Physical Review B.
\bibitem{kim99} C. S. Kim, A. M. Satanin, Y. S. Joe, R. M. Cosby, Phys. Rev. B 60 (1999) 10962.
\bibitem{ladron06} M. L. Ladr\'{o}n de Guevara, P. A. Orellana, Phys. Rev. B 73 (2006) 205303.
\bibitem{voo05} Khee-Kyum Voo, C. S. Chu, Phys. Rev. B 72 (2005) 165307.
\bibitem{lee06} M. Lee, C. Bruder, Phys. Rev. B 73 (2006) 085315.
\bibitem{wang06} R. Wang, J. -Q. Liang, Phys. Rev. B 74 (2006) 144302.
\bibitem{bag93} E. Tekman, P. F. Bagwell, Phys. Rev. B 48 (1993) 2553.
\bibitem{south83} B. W. Southern, A. A, Kumar, J. A. Ashraff, Phys. Rev. B 28 (1983) 1785.
\bibitem{stone81} A. Douglas Stone, J. D. Joannopoulos, D. J. Chadi, Phys. Rev. B 24 (1981) 5583.
\bibitem{swarn04} S. Bandopadhyay, B. Dutta-Roy, H. S. Mani, Am. J. Phys. 72 (2004) 1501.
\end{itemize}
%\end{thebibliography}   
\end{document}